\documentclass[manuscript]{aastex}
\usepackage{graphicx}
\usepackage{epsfig}
\usepackage{amsmath}
\usepackage{lineno}
\usepackage{comment}

\begin{document}
\shorttitle{Coupling of Convective Flows and Mangetic Field}
\shortauthors{Fang et al.}
\title{Dynamic Coupling of Convective Flows and Magnetic Field during Flux Emergence}

\author{Fang Fang\altaffilmark{1}, Ward Manchester IV\altaffilmark{1}, 
  William P. Abbett\altaffilmark{2}, 
  Bart van der Holst\altaffilmark{1} }
\affil{\altaffilmark{1} Department of Atmospheric, Oceanic and Space Sciences, 
  University of Michigan, Ann Arbor, MI 48109, USA}
\affil{\altaffilmark{2} Space Sciences Laboratory, University of California, 
  Berkeley, CA 94720, USA}


\begin{abstract}
We simulate the buoyant rise of a magnetic flux rope from the solar convection
zone into the corona to better understand the energetic coupling of the solar
interior to the corona. 
The magnetohydrodynamic model addresses the physics of radiative cooling, 
coronal heating and ionization, which allow us to produce 
a more realistic model of the solar atmosphere. 
The simulation illustrates the process by which magnetic flux emerges 
at the photosphere and coalesces to form two large concentrations of 
opposite polarities.
We find that the large-scale convective motion in the convection zone 
is critical to form and maintain sunspots, 
while the horizontal converging flows in the near surface layer 
prevent the concentrated polarities from separating. 
The foot points of the sunspots in the convection  zone exhibit
a coherent rotation motion, resulting in the increasing helicity of the coronal
field. Here, the local configuration of the convection causes the convergence
of opposite polarities of magnetic flux with a shearing flow along the polarity
inversion line.  During the rising of the flux rope, 
the magnetic energy is first injected through the photosphere by the emergence, 
followed by energy transport by horizontal flows, 
after which the energy is subducted back to the convection zone by 
the submerging flows.

\end{abstract}





\section{Introduction} \label{intro}


The solar magnetic field emerges at the surface over a wide range of scales, from 
ubiquitous ephemeral regions as small as $10^{16}$ Maxwells, to active regions
as large as $10^{23}$ Maxwells, which emerge at low to midlatitudes  
\citep[]{hagenaar2003,parnell2009}. 
Observations from satellites and ground-based telescopes 
find that the configuration of the magnetic field plays an important role in 
the solar eruptive events, i.e., Coronal Mass Ejections (CME) and solar flares 
\citep[]{canfield1999}. 
It is thus of great importance to study the building up of the magnetic field
in the solar atmosphere, as it rises from the convection zone.
However, the study of the transport of magnetic flux and energy
has been hampered by the invisibility of subsurface structures. 
With the time-distance helioseismic analysis, \cite{sasha1996} and \cite{zhao2003} 
provide a view on the horizontal and vertical flow velocities on the subsurface layers 
under sunspots and identify shear flows and rotation of the sunspots 
underneath the surface, which may build up the energy and helicity in the atmosphere. 

Over the past decades, the development of numerical models have also greatly  
improved our understanding of the dynamics and energetics of magnetic flux emergence. 
\cite{shibata1989} describes a two-dimensional magnetohydrodynamic 
(MHD) simulation on the emergence of a horizontal magnetic flux rope from the 
photosphere into the chromosphere using a two-layered atmosphere. 
\cite{fan2008} and \cite{jouve2009} carry out sets of anelastic MHD simulations 
on the buoyant rise of the magnetic flux tube from the base of convection zone 
to near the top, respectively. 
In particular, \cite{fan2008} shows the rotation of the flux tube driven 
by the Lorentz force at the two ends while the twisted tube is bent. 
\cite{manchester2000} suggested that shearing motion driven by the Lorentz force
draws the magnetic field parallel to the Polarity Inversion Line (PIL), 
which was demonstrated in simulations of emerging flux ropes
\citep[]{fan2001,magara2003,abbett2003,manchester2004,archontis2008,mactaggart2009}.
\cite{magara2003} found that during emergence, energy flux through the photosphere
is first dominated by the vertical flows 
while horizontal flows dominate the later phase.
The energy transport of shear flows naturally provides an energy source for CMEs
\citep[]{manchester2007,manchester2008}.
Simulations have also revealed that shear flows driven by the Lorentz force 
can produce eruptions in both magnetic arcades \citep[]{manchester2003} and 
emerging flux ropes \citep[]{manchester2004,archontis2008,mactaggart2009} 
providing further evidence of a mechanism for CMEs, flares and filament eruptions.

With the availability of more computational resources, numerical models are able to 
take into account the thermodynamic processes in the solar interior and atmosphere, 
and produce a realistic convection zone overlaid by a self-consistent upper atmosphere 
\citep[]{vogler2005,stein2006,hansteen2006,abbett2007,gudiksen2011}. 
Such realistic simulations reveal a wealth 
of complex dynamic interactions between the magnetic field and the convective flows. 
\cite{cheung2007,tortosa2009} study the rise of buoyant magnetic flux tubes from the 
near-surface convection zone into the photosphere and chromosphere and find the 
fundamental role of convective flows in the emergence of the magnetic flux at the 
photosphere.
\cite{cheung2007,martinez2008,martinez2009} 
simulate an emerging flux tube, which increases the size of the photospheric granules 
and alters the chromospheric structure. More discussion on the interaction between the 
magnetic field and the convective flows can be found in \cite{fan2004,nordlund2009}.

Radiative MHD simulations on a larger scale \citep[]{rempel2009,cheung2010} report
the formation of the sunspots and active regions. \cite{rempel2009,rempel2011} find
outflows in the penumbral structure driven by the Lorentz forces at the surface and 
by the convective flows in the deeper layers. 
\cite{cheung2010} simulates the formation of a pair of sunspots in an active region 
with a magnetic semi-torus advected through the bottom of the domain 
and finds the mass removal in the magnetic flux driven by the reconnection, 
and the migration of the flux due to horizontal flows. 
With the inclusion of turbulent convection \citep[]{fang2010}, it is shown that 
the transport of energy into the corona becomes even more critical as downdrafts 
return much of the magnetic energy back below the photosphere. This simulation 
treats the emergence of a weak axial flux of $3.3\times10^{19}$ Mx, 
which is quickly shredded and dispersed to the intergranular lanes 
by the convection motions.  

These radiative MHD simulations illustrate the importance of turbulent convective flows 
in the emergence of the magnetic structures.  In light of these 
previous results, we expand upon the work of \cite{fang2010} by simulating the 
emergence of a larger twisted magnetic flux rope emerging from deeper in the 
convection zone, and study the energetics during its rise through the turbulent 
plasma to reside in the hot corona. 
With this work, we strive to more deeply understand the energetic coupling between
the convection zone and the corona provided by the magnetic field. 
In the following sections, we present the results of our simulation. 
Section \ref{method} describes the numerical methods and the simulation steps. 
In Section \ref{results}, we study the results by analyzing the evolution of the 
magnetic and energetic flux under the impact of the convective flows. 
Finally, we discuss the conclusions of our simulation and its implication on 
further simulations. 


\section{Numerical methods} \label{method}
For our simulation, the MHD equations are solved with the Block-Adaptive Tree 
Solar-wind Roe Upwind Scheme (BATS-R-US; \cite{powell1999}). 
Additional energy source terms are implemented to include the thermodynamic 
processes in the solar atmosphere, i.e. surface cooling at the photosphere, 
radiative cooling of the corona and magnetic flux-related heating 
in the corona \citep[]{abbett2007}. 
We describe the radiative losses in the low-density corona 
as optically thin radiation and artificially extend the cooling function to 
treat the solar surface temperature as well. 
The coronal heating is approximated by \cite{pevtsov2003}, which gives an empirical 
relationship between the heating rate and the unsigned magnetic flux 
at the photosphere.
Our model neglects electron thermal conduction along magnetic field lines, 
since the simulations here focus on the transfer of energy and magnetic field from 
the convection zone to the corona. 
The application of a tabular, non-ideal equation of state takes account of the 
ionization and the excitation of particles in the dense and hot solar 
interior \cite[]{rogers2000}. 
Horizontal boundary conditions are periodic, and the lower boundary condition 
sets the density and temperature of the same values found in the initial state
of the atmosphere, while keeping the vertical momentum reflective across 
the boundary. The upper boundary is closed. 
The details of the model are described in \cite{fang2010}. 


\subsection{Background Atmosphere} \label{atm}
To develop the coronal model, we first extract the averaged state of the 
solar atmosphere from the model of \cite{fang2010}, 
at a depth of 2.5 Mm below the photosphere. 
In the convection zone, the plasma is adiabatic as the heating source is 
absent and radiative cooling is negligible in the optically thick medium.
The entropy is then nearly invariant throughout the deep convection zone. 
Here the deep convection zone is defined as deep in the simulation domain, 
which covers the upper 10\% (20 Mm) depth of the solar convection zone. 
Under the assumption that the plasma is both isentropic and in hydrodynamic 
equilibrium, we solve the following equations for the structure 
of the atmosphere in the deep convection zone with the extracted values 
providing the upper boundary condition: 

\begin{equation}
  \label{force}
  \nabla p = \rho {\bf g},
\end{equation}
\begin{equation}
  \label{isoentropy}
  log(\frac{p}{\rho^{\gamma}}) = C,
\end{equation}
where $p$, $\rho$ and ${\bf g}$ are the plasma pressure, 
mass density and gravitational acceleration. C is the  
average entropy value in the convection zone at $z = -2.5$ Mm. 
$\gamma$ is the adiabatic index, which can be obtained 
from the tabular equation of state using the pressure and density values, $p$ 
and $\rho$. For the purpose of our simulation here, 
the atmosphere is extended to 21 Mm below
the photosphere by solving Equations \ref{force} and \ref{isoentropy}. 
The stratification of the density and pressure of the extended atmosphere 
in the convection zone are found to be in agreement with that of \cite{stein2011}

This solution is then used as an initial condition for a one-dimensional 
calculation in which we apply the surface cooling.  The plasma in the 
simulation domain then goes through an hour of thermal relaxation to  
form a supraadiabatic background atmosphere, which is unstable to 
convective motion.  The next step is to perform the full 3-D
simulation with the size of the domain set to $30\times30\times42$ Mm$^{3}$, 
extending 21 Mm below the surface and 21 Mm above in the vertical directions,
while the horizontal area of the domain is comparable to the size of 
a small active region.  In the supraadiabatically stratified atmosphere, 
convective motion starts immediately after a small energy perturbation, 
and relaxes over a period of roughly one turnover time of the convection
extending to a depth of 21 Mm.  After the convective motion has relaxed, 
we impose an initial vertical magnetic field $B_{z}$ of 1 G to the domain, 
and turn on the heating and radiative cooling term in the coronal region. 
With the magnetic-flux-related coronal heating term \citep[]{pevtsov2003}, 
the coronal temperature is heated up to $1\times10^{6}$ K within 2 hours. 

\begin{figure*}[ht!]
  \begin{minipage}[t] {1.0\linewidth}
    \begin{center}
      \includegraphics[width=150mm]{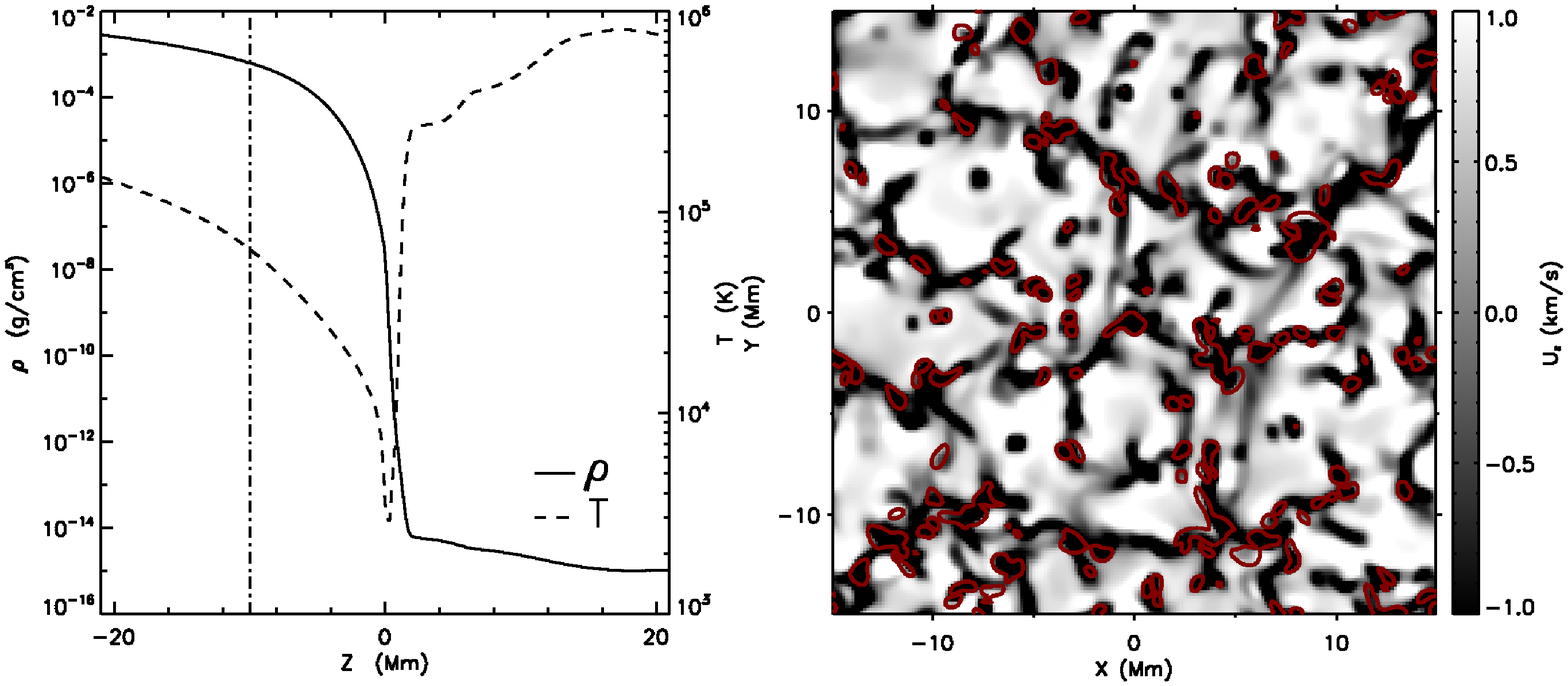}
    \end{center}
  \end{minipage}\hfill
  \caption{Left: The vertical stratification of the density (solid) and the 
    temperature (dashed) of the solar atmosphere. The dash-dotted line 
    indicates the initial location of the inserted flux rope.
    Right: The vertical velocity structure at $z = -3$ Mm. 
    Red lines show regions with $|B_{z}|$ greater than 10 G.}
  \label{initatm}
\end{figure*}

The left panel in Figure \ref{initatm} shows the averaged vertical 
stratification of the density and temperature in the simulation domain. 
The density drops by about 5 orders of magnitude from the bottom of the 
domain to the photosphere, and the temperature drops by 2 orders of magnitude. 
Above the photosphere, the density drops by another 7 orders of magnitude, 
and the temperature increases to 1 MK within 2 Mm.
The strong variation in the atmospheric stratification puts large 
requirements on spatial resolution and restrictions on the 
explicit time step, given the high temperatures and subsequent sound speeds. 
However, the application of the Adaptive Mesh Refinement in 
BATSRUS greatly facilitates the simulation in saving computational 
resources while keeping the necessary refinements to resolve the local 
turbulent structures. The grid size in our simulations has three levels of 
refinement with cubic cells of size 37.5 km, 75 km,  and 150 km.  
The grid is refined in horizontal layers with higher resolution near the surface  
where temperature and density gradients are largest.

The right panel in Figure \ref{initatm} shows the 
variation of the vertical velocity $u_{z}$ at $z = -3$ Mm. Red lines outline
the regions with concentrations of $B_{z}$ field, which coincides well with 
the down-flowing area. The right panel in Figure \ref{initrope} illustrates 
the structure of $u_{z}$ on a vertical cut of the convection zone.
Small-scale down-flowing plasma merges and forms the large-scale persistent 
downdrafts in the deep convection zone. The large-scale downdrafts are very
important in the formation of pores during the building up 
of the active region, discussed in Section \ref{dipoles}.  

\subsection{Flux Rope} \label{rope}
The stratification of the atmosphere in the simulation domain is maintained 
self-consistently with the implementation of the thermodynamic processes 
in the model. The superadiabatic stratification provides an unstable 
background atmosphere for turbulent convective motion.  As shown by 
\cite{cheung2010}, convective motion plays an important role in the formation
of the sunspots and active region.  However, the role of the turbulent 
convection in the emergence of the magnetic flux has not been clearly 
understood yet, particularly at larger scales.  The aim of our simulation 
here is thus to study the transfer of the energy and magnetic flux of 
the rope during its rise from the deep convection zone, and its 
interaction with the surrounding turbulent medium. 
So after the generation of the convection zone with a hot corona, 
we linearly superimpose a flux rope upon the ambient magnetic field
in the deep convection zone, at $z = -10$ Mm, 
indicated by the dot-dashed line in the left panel of Figure \ref{initatm}.
The initial flux rope is centrally buoyant and twisted along the $x -$ axis, 
as described in \cite{fan2001} and \cite{manchester2004} 
by the following equations:

\begin{equation}
  \label{brope}
        {\bf B} = {\bf B}_{\mathrm{amb}} + B_{0}e^{-r^{2}/a^{2}}\hat{\boldsymbol x}
        + qrB_{0}e^{-r^{2}/a^{2}}\hat{\boldsymbol\theta},
\end{equation}
\begin{equation}
  \label{densitypert}
  \rho = \rho_{0}(1-\eta e^{-x^{2}/\lambda^{2}}),
\end{equation}
\begin{equation}
  \eta = 
  \frac{\frac{1}{2}\left[B_{0}e^{-r^{2}/a^{2}}\right]
    ^{2}\left[-1+\frac{1}{2}q^{2}(1-\frac{2r^{2}}{a^{2}})\right]}
       {p_{0}},
\end{equation}
\begin{equation}
  p = p_{0}(1-\eta).
\end{equation}
Here $r$ is the distance from the axis of the rope, 
and $a = 1$ Mm, is the radius of Gaussian decay. 
$q = -1.5$, is the twisting factor. 
$\lambda = 6$ Mm, is the length of the buoyant section. 
$p_{0}$ is the thermal pressure without the inserted flux rope.
$B_{0} = 50$ kG, is the strength of the magnetic field at the axis of the rope,
giving a minimum plasma beta value of 50. 
${\bf B}_{\mathrm{amb}}$ is the ambient magnetic field. 
The total axial flux on the cross section of the initial flux rope 
is $1.5\times10^{21}$ Mx. 
Figure \ref{initrope} shows the initial flux rope embedded with 
the convective plasma. The left panel illustrates 3-D isosurfaces of 
the large scale convective flows around the flux rope 
in the deep convection zone, while the right panel shows the structure of 
$u_{z}$ in the convection zone on the $x-z$ plane with the flux rope. 
The complexity of the convective flows around the flux rope has a strong 
influence on the emerging process from the deep convection zone with 
large-scale downflows.


\section{Results} \label{results}

\begin{figure*}[ht!]
  \begin{minipage}[t] {1.0\linewidth}
    \begin{center}
      \includegraphics[width=160mm]{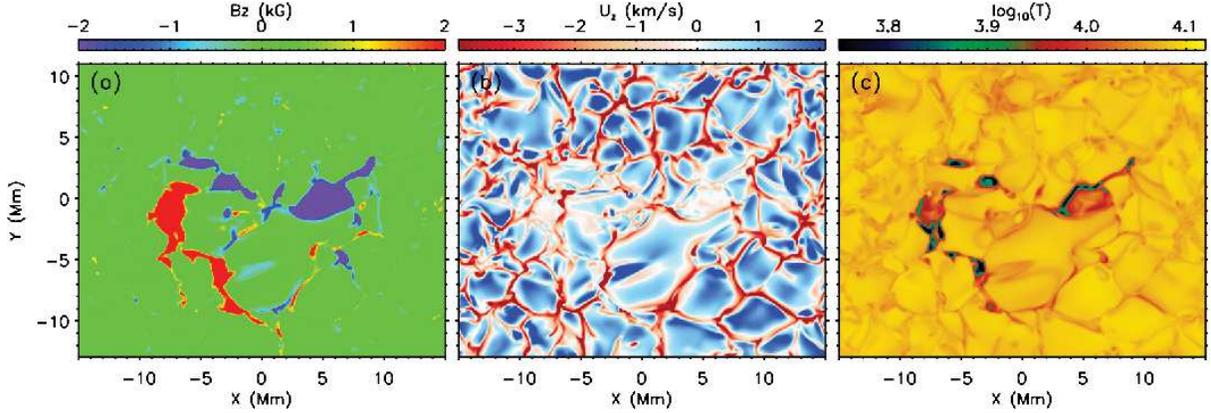}
    \end{center}
  \end{minipage}\hfill
  \caption{The structure of $B_{z}$ (a), $u_{z}$ (b) and $T$ (c)
    at $z = -1$ Mm plane at $t =$ 5:09:00.}
    \label{bzuzt}
\end{figure*}

The flux rope is initially embedded in the deep convection zone, where it
is buffeted by convective down- and up- flows. On the way from  $z = -10$ Mm to 
the photosphere, 
the turbulent flows of various scales 
significantly reshape and redistribute the magnetic flux.  On the photosphere, 
the magnetic flux first emerges as many small-area bipoles, which then sort 
themselves to form two large concentrations of opposite polarities via coalescence. 
Figure \ref{bzuzt} shows the structure of the small active region 
at $z = -1$ Mm. It contains two pores of opposite magnetic polarities, with 
smaller-size granules and cooler temperature. 
The formation of pores by the process
of coalescence at the photosphere is revealed by both remote observations and 
previous numerical simulations \citep[]{vrabec1974,zwaan1985,cheung2010}.  
However, the mechanism that 
transfers the magnetic flux from the tachocline, which is believed to be the 
origin of the active region magnetic field, to the solar surface is still to be 
determined. To date, global scale simulations of the solar interior can not yet 
resolve individual flux tubes.  Here, we examine the evolution of the magnetic flux rope 
in the deep convection zone and the photospheric and coronal response to its 
emergence.

\subsection{Formation of Pores} \label{dipoles}
\begin{figure*}[ht!]
  \begin{minipage}[t] {1.0\linewidth}
    \begin{center}
      \includegraphics[width=155mm]{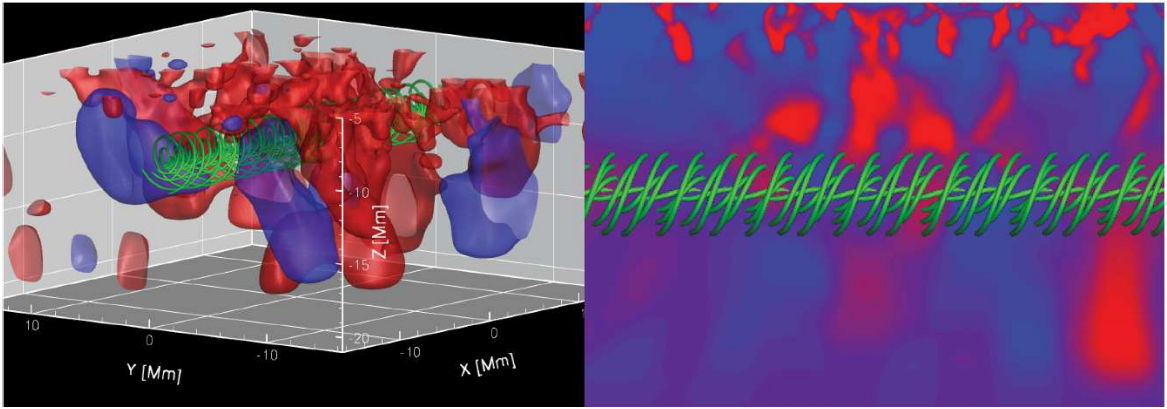}
    \end{center}
  \end{minipage}\hfill
  \caption{The initial structure of magnetic flux rope with color showing 
    the vertical velocity in the convection zone.
    ~~Left: magnetic flux rope embedded in the convective flows with 
    blue isosurfaces indicating upflows at 1 km~s$^{-1}$ and red downflows 
    at -1 km~s$^{-1}$.
    ~~Right: $x-z$ plane of the convection zone with the inserted flux rope.}
  \label{initrope}
\end{figure*}

\begin{figure*}[ht!]
  \begin{minipage}[t] {1.0\linewidth}
    \begin{center}
      \includegraphics[width=160mm]{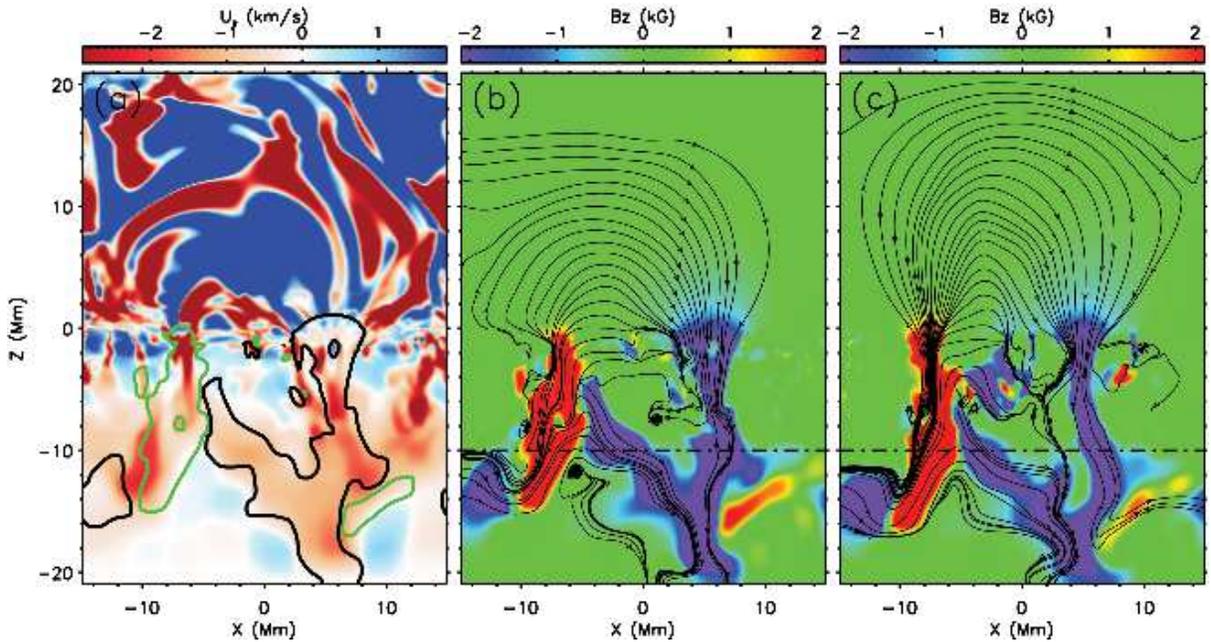}
    \end{center}
  \end{minipage}\hfill
  \caption{The structure of $u_{z}$ (a), $B_{z}$ (b) and $B_{z}$ (c)
    in the $y = 0$ plane at $t =$ 4:50:00 (a and b) and $t =$ 5:22:00 (c).
    The black and green lines in Panel (a)
    outline $B_{z} = $ -1, 1 kG, respectively. 
    The dash-dotted line in Panel (b) and (c) indicate the initial location 
    of the axis of the inserted flux rope.
    And the black lines in Panel (b) and (c) indicate 
    the magnetic field lines by ignoring the $B_{y}$ component to show
    the direction of the field confined to the $x-z$ plane. }
  \label{y=0}
\end{figure*}

The horizontal flux rope at $z = -10$ Mm, shown by Figure \ref{initrope}, 
rises in the central section due to the depletion of the density and 
upwelling convective flows. 
However, the two ends of the central section are embedded in 
large-scale downflows present in the convection zone when the flux rope 
is initiated at $t = 0:00:00$. 
The downflows are illustrated by the blue isosurfaces and color contours in 
Figure \ref{initrope}. Panel (a) of Figure \ref{y=0} 
shows the structure of $u_{z}$ on the $x-z$ plane at $t = 4:50:00$, 
when the downflows are still present in the convection zone at the two 
ends of the emerged flux rope. 
The region of the convective downflows in Panel (a) of Figure \ref{y=0} 
appears in great accordance with the black and green contour lines indicating 
the polarity of the pores. 
This consistency between the down flowing and magnetic concentrated regions 
suggests a causal relationship between the formation of the magnetic pores and 
the large-scale downflows.
Figure \ref{dipole} illustrates the temporal evolution of the 3-D magnetic field, 
colored by the $u_{z}$ value of the local plasma during the rising of the 
central section of the flux rope. The long-lasting, large-scale downflow, 
indicated by the blue color, drags down the two endpoints of the rising part and
fixes them in the deep convection zone, forming an $\Omega$-shape emerged 
flux rope within 2.5 hours. The downflow maintains the pores and 
prevents the two pores of opposite polarities from separation or drifting apart, 
while the central section of the flux rope 
emerges and expands in the upper domain. Thus with the two foot points deeply 
embedded and fixed in the convection zone, the emerged magnetic flux remains 
highly concentrated in a relatively small area at the photosphere. 

In Panel (b) of Figure \ref{bzuzt} and Panel (a) of Figure \ref{y=0}, 
small-scale convective granules in the near surface layers appear 
inside each of the pores. 
\cite{emonet2001} also simulates the diminishing horizontal scale 
of the granules with increasing vertical magnetic field.
While the magnetic field modifies the convective granules,
the downward flows produce a bulb of colder plasma. 
The pressure imbalance with the surrounding material thus causes the 
flux tube to collapse and increases the strength of the magnetic field. 
The flux tube approaches equilibrium again with higher magnetic pressure balancing 
the surrounding material. 
With the convective collapse process 
\citep[]{parker1978,spruit1979,nagata2008}, 
the magnetic field strength at the surface can be increased up to 4 kG, 
much higher than the equipartition field strength with the kinetic 
energy of the surrounding plasma. 
These near-surface processes, due to the convective motion, play a very important 
role in reorganizing the magnetic flux after its emergence at the photosphere.
The magnetic field of the bipolar structures at the photosphere is 
intensified by the convective collapse in the near surface layers, 
which is shown by Panel (c) of Figure \ref{y=0}. 

\subsection{Rotation of the Pores} \label{rotation}

The magnetic flux rope travels through the 10 Mm distance and 
approaches the photosphere after 2.5 hours. Figure \ref{bzz=0} shows the evolution
of the structure of the $B_{z}$ field at the photosphere, with arrows representing 
the horizontal velocity field. 
The magnetic field concentrates as narrow bands of 
bipolar fluxes, as shown in Panel (a) and (b), in the regions between 
the major pores. The horizontal velocity, represented by the white arrows, 
reveals that the bipolar fluxes are moving in opposite directions, 
toward the major pores of the same polarity. 
For example, in Panel (a), the rectangle outlines a region with flux emergence, 
where the negative flux is moving in the upper right direction 
into the negative pores and the positive flux bands are moving 
in the lower left direction into the positive pores. 
The coalescence of the 
small-scale fluxes into the major pores facilitates the accumulation of the
magnetic flux on the surface and, therefore, the formation of the large pores 
shown in Panel (c) and (d). \cite{cheung2010} simulates the formation of 
an active region and finds that the counterstreaming motion of opposite
polarities is driven by the Lorentz force.

\begin{figure*}[ht!]
  \begin{minipage}[t] {1.0\linewidth}
    \begin{center}
      \includegraphics[width=140mm]{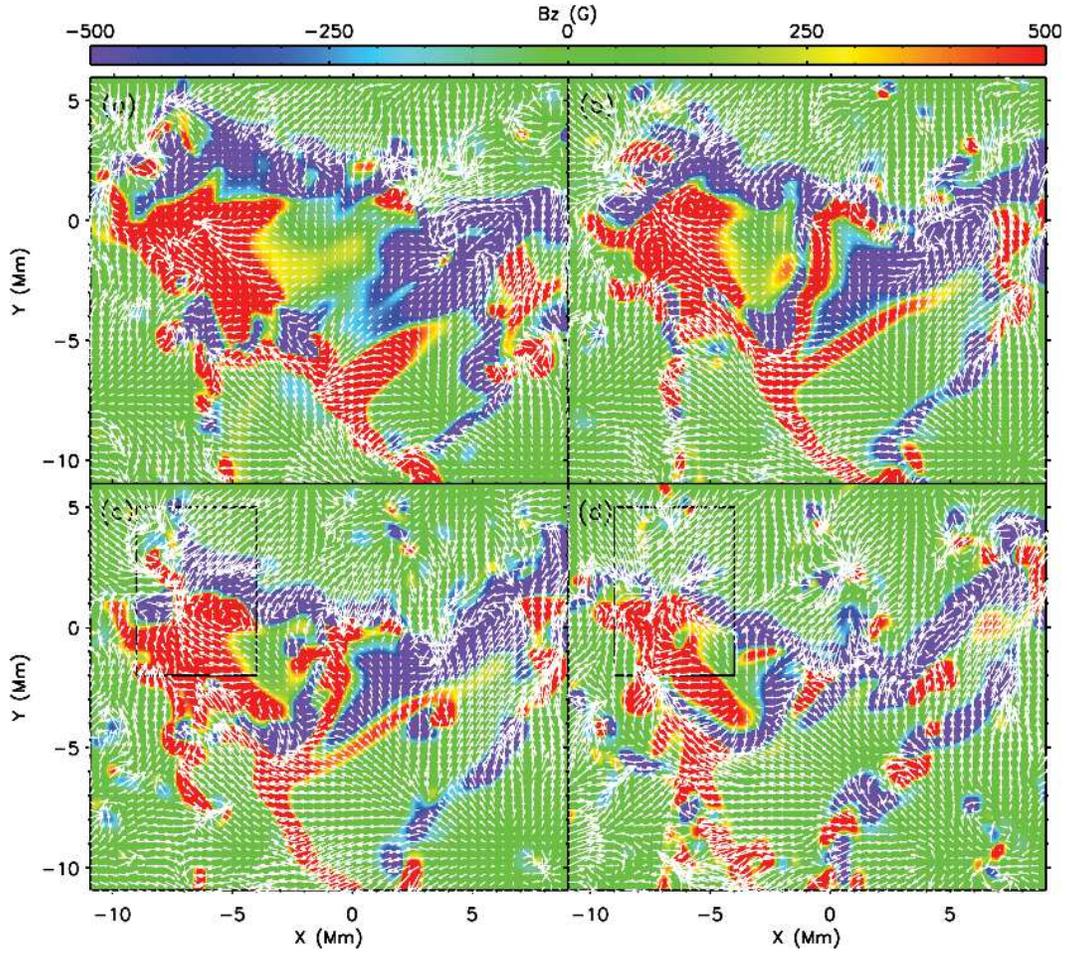}
    \end{center}
  \end{minipage}\hfill
  \caption{The structure of $B_{z}$ field at $z = -3$ Mm at t = 4:40:00 
    (a), 5:00:00 (b), 5:10:00 (c), and 5:30:00 (d). The arrows show 
    the horizontal velocity field.}
  \label{bzz=-3}
\end{figure*}

\begin{figure*}[ht!]
  \begin{minipage}[t] {1.0\linewidth}
    \begin{center}
      \includegraphics[width=140mm]{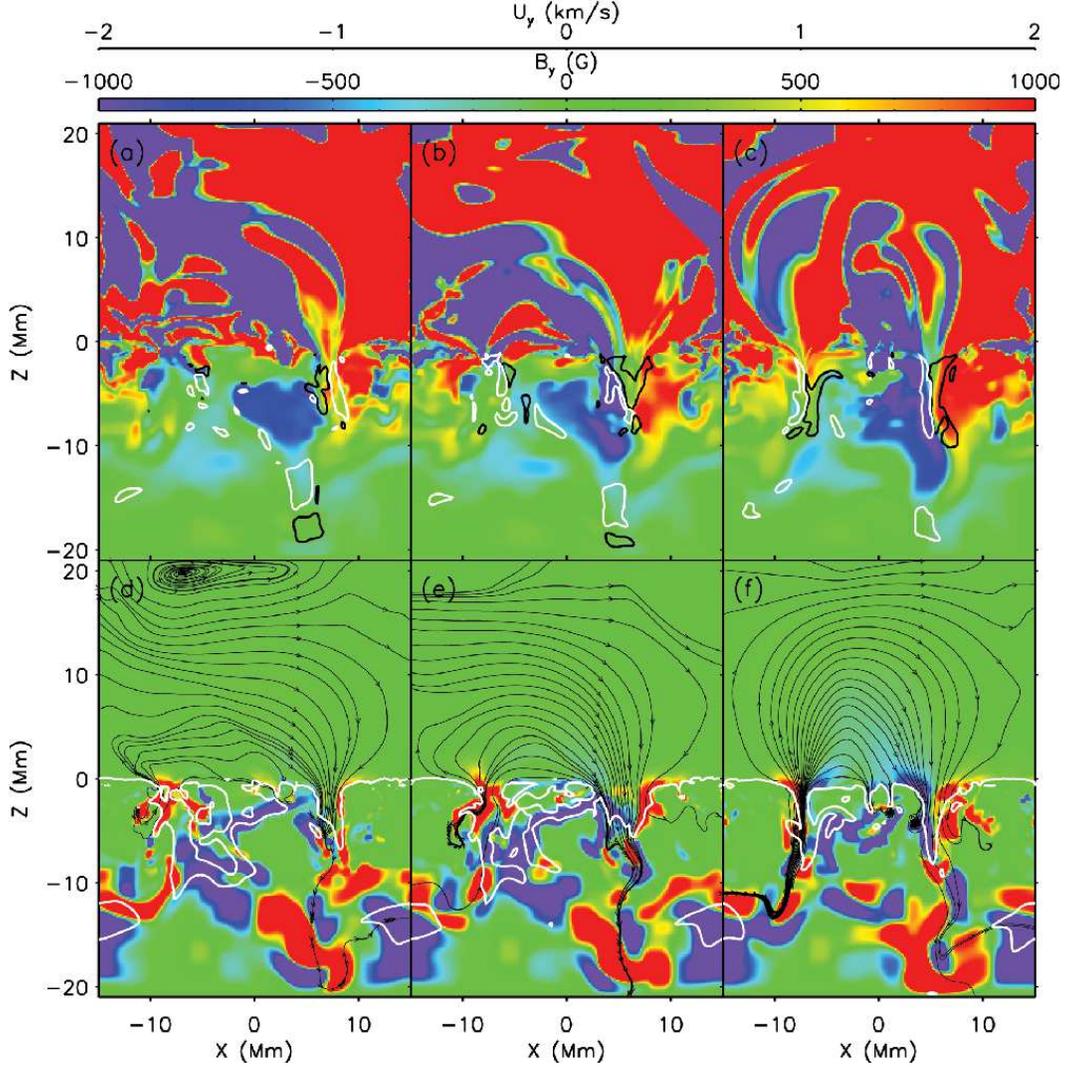}
    \end{center}
  \end{minipage}\hfill
  \caption{The structure of $u_{y}$ (upper panels) and $B_{y}$ (lower panels) 
    at time $t =$ 4:21:00 (a and d), 4:37:00 (b end e) and 5:13:00 (c and f) 
    in the $y = 0$ plane.
    The black and white contour lines in the upper panels  
    represent regions with Lorentz force density $f_{y} =$ -20 and 20 dyne~cm$^{3}$ respectively.
    And the black lines with arrows in the lower panels 
    indicate streamlines of ($B_{x}, B_{z}$) while white lines outline 
    Alfv\`enic Mach number $u/u_{A} =$ 1.}
  \label{roty=0}
\end{figure*}

The large pores of negative polarity show a coherent pattern of rotation 
at the photosphere after their formation in Panel (a) of Figure \ref{bzz=0}. 
The rotation of the pores persists during the emergence and the increase of 
magnetic flux at the photosphere. However, the positive polarity on the left 
does not present a complete rotation pattern during the emerging phase, 
but the rotation is interrupted by the horizontal motions of the convective flow. 
The coherent rotation starts to develop on the positive pore after 5.5 hrs,
shown by Panel (e) and (f) in Figure \ref{bzz=0}. 
Figure \ref{bzz=-3} illustrates the evolution of $B_{z}$ and the horizontal 
velocity fields at $z = -3$ Mm, in the convection zone. 
Here, we observe a coherent rotation at this depth on the negative polarity as well. 
The question is then, to what depth does the rotation extend. Thus we examine
the structure of $u_{y}$ on the $y=0$ plane during the rising of the flux rope 
in the upper panels in Figure \ref{roty=0}. 
The reversal of the direction of $u_{y}$ in the right side of the domain 
corresponds to the projection of the rotation 
of the negative polarity on the $y = 0$ plane. The coherent rotation starts to 
extend downward at $t = 4:00:00$ and approaches the depth of -10 Mm in 21 mins. 
(see Panel (a)). Panel (c) shows a very coherent pattern of rotation on the negative
polarity at $t = 5:13:00$, while on the positive pore on the left, the rotation is 
not obvious. 

Sunspot rotation has long been observed and studied in detail and has been
found in association with CMEs \citep[]{brown2003,kazachenko2009}.
The rotation mechanism for sunspots found at work in our simulation was 
first described in \cite{parker1979}. 
\cite{longcope2000} simulates the increase of the helicity of coronal field 
due to the rotation of the photospheric footpoints in a dynamic model.
Ideal MHD simulations by \cite{fan2009} illstrate the rotation driven by 
torsional Alfv\`en waves and the development of sigmoid-shaped field lines
during the flux emergence from the top layer of the convection zone.
\cite{brown2003} and \cite{min2009} study the behavior of 
the rotating sunspots in solar active regions 
and found that the rotation speed varies with time, 
radius and angular spacing, with an increase in rotation speed in the penumbra. 
\cite{sasha1996,zhao2003} find twists of the magnetic field and 
vortical flows around the sunspots in the convection zone underneath 
a rapidly rotating sunspot area.
The flows underneath the negative pore in our simulation 
also exhibit a rotating pattern extending 10 Mm down into the convection zone.
The question is then, what dynamic mechanism during flux emergence 
causes the observed variation in the rotation speed.
Our simulation enables a detailed investigation of how sunspot rotation develops
in the complex circumstances of the convection zone.

The development of the coherent rotation is accompanied by the presence 
of a strong Lorentz force in the azimuthal direction 
(see the upper panels in Figure \ref{roty=0}), which is defined as 
$f_{y} = j_{z}B_{x} - j_{x}B_{z}$.
The black and white contour lines in 
the upper panels of Figure \ref{roty=0} represent areas with 
strong negative and positive $f_{y}$ respectively.
We find that the Lorentz force is driving the coherent rotation 
as a torsional Alfv\`en wave. At time $t = 4:21:00$, 
shown by Panel (a), the Lorentz force is in the same direction as $u_{y}$, 
thus acting to accelerate the rotation. 
However, the Lorentz force reverses direction at time $t = 4:37:00$ 
(see Panel (b)) and decelerates the rotation. 
Panel (c) reveals the structure of the Lorentz force at time $t = 05:13:00$, 
which runs in the opposite direction of the rotation. 
The flux rope has rotated past its equilibrium point, 
and the Lorentz force reverses with the gradient 
of the azimuthal ($B_{y}$) component of the field as the flux rope further emerges. 
The rotation of the pores untwists the field in the convection zone and produces 
a twisted magnetic field in the corona, as shown by 
the lower panels of Figure \ref{roty=0},
which illustrates the $B_{y}$ field on the $y = 0$ plane. As such, the rotation also 
provides a mechanism to build up the helicity and magnetic energy in the 
coronal magnetic field. The white lines in the lower panels outlines where 
the Alfv\'enic Mach number $u/u_{A} = 1$. The consistency between the structures
of the Alfv\'enic Mach number and the Lorentz force $f_{y}$ suggests that in the pore
region the magnetic field dominates over the plasma motion, and is responsible for 
the rotating flows.

In Figures \ref{bzz=0} and \ref{bzz=-3}, the positive polarity does not present a 
coherent rotation as the negative polarity. However, in Panel (c) and (d) of Figure 
\ref{bzz=0}, we observe a strong flux cancellation in the region outlined by the 
rectangles from time $t = 5:00:00$ to $t = 5:35:00$. 
\cite{cheung2010} reports that magnetic reconnection takes place within
the U-loops formed by the convective downflows, which can remove the unsigned 
magnetic flux on the photosphere.
In our simulation, the total amount of the magnetic flux that is cancelled is 
10\% of the total unsigned flux on the photosphere. 
The flux cancellation strongly interrupts the 
coherency of the rotation on the positive polarity. The horizontal velocity fields 
exhibit a converging flow at the two opposite polarities of the cancelled flux, both 
at the photosphere and in the convection zone, yielding a high gradient in the 
magnetic field. 
The close examination of the flux cancellation event will be the topic of a future paper. 
Another interesting feature we observe in our simulation is that 
the horizontal velocity field in Figure \ref{bzz=-3} shows a converging horizontal 
flow field around the magnetic flux region. This converging flow constrains 
the total area of the magnetic concentration, and thus prevents the magnetic 
pores from expansion and separation. 

\subsection{Energy Fluxes} \label{fluxes}
\begin{figure*}[ht!]
  \begin{minipage}[t] {1.0\linewidth}
    \begin{center}
      \includegraphics[width=160mm]{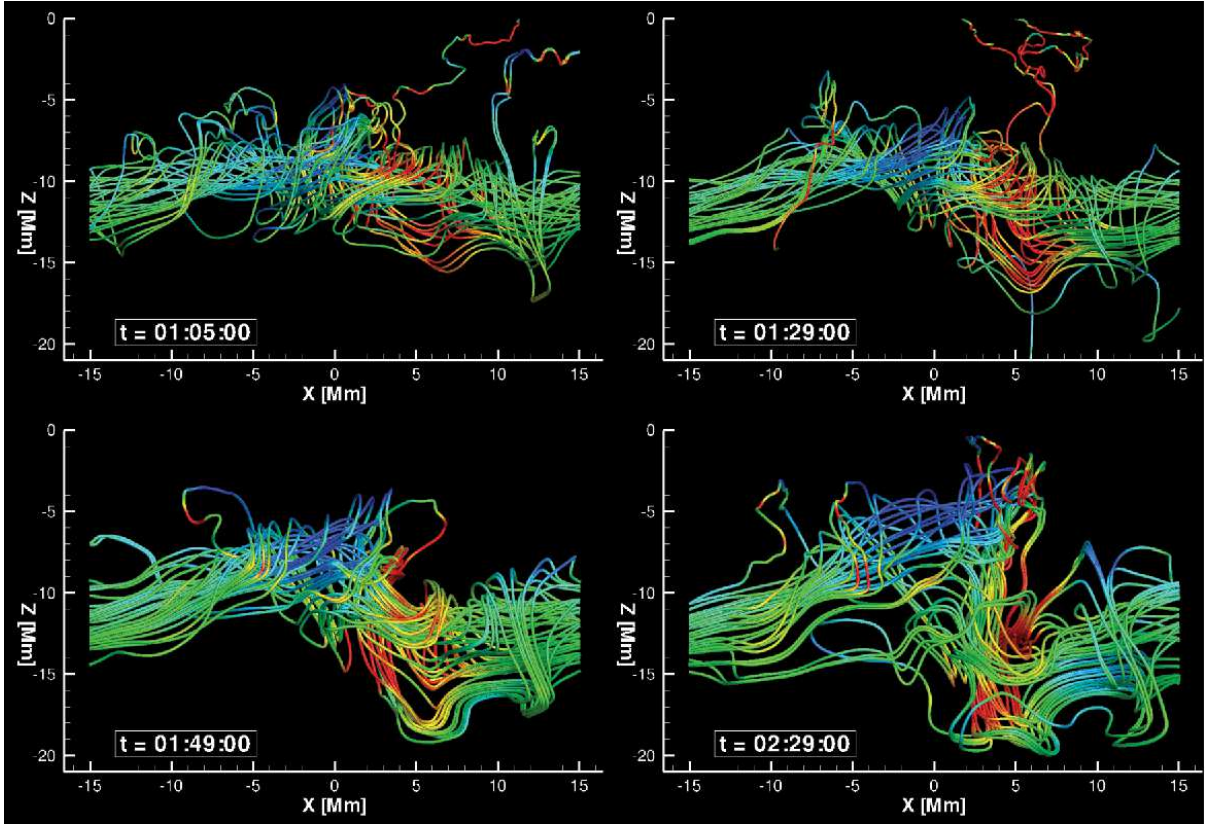}
    \end{center}
  \end{minipage}\hfill
  \caption{The temporal evolution of 3-D magnetic field lines 
    colored by local $u_{z}$ values with red indicating downflows 
    and blue upflows from -2 km~s$^{-1}$ to 1 km~s$^{-1}$, 
    at $t =$ 1:05:00, 1:29:00, 1:49:00 and 2:29:00.}
  \label{dipole}
\end{figure*}

The magnetic flux rope approaches the photosphere within 2.5 hours after 
its initialization, and the photospheric magnetic flux, i.e. the emerged flux, 
reaches its maximum value at time $t = 5:11:00$. Afterward, the magnetic pores 
start to decay slowly for the rest of the simulation.  However, even at its 
maximum emergence, the magnetic flux rope is far from being fully above the 
photosphere. This is clearly shown by Panel (b) and (c) of
Figure \ref{y=0}, where the negative polarity is split into two parts, with one 
part emerging and forming pores while the other is sinking to the deep convection 
zone.  The lower right panel of Figure \ref{dipole} shows the 3-D magnetic field 
lines of the flux rope when it first approaches the photosphere.  The red contour 
color over the field lines represents upflows while blue represents downflows.
Only about half of the original flux rope is rising with upflows, 
while the other half remains almost stationary or sinks in the convection zone. 

It is of great interest to study the mechanism for transporting the initial 
magnetic flux from the deep convection zone into the photosphere, where the 
magnetic pores appear, and further into the coronal region.  Here we calculate 
the temporal evolution of the total unsigned magnetic flux at the photosphere 
$(z = 0)$, $z = -3$ Mm and $z = 3$ Mm, shown by the upper panel of Figure \ref{flux}. 
At time $t = 4:49:00$, the magnetic flux at $z = -3$ Mm reaches its maximum value of 
$2.63\times10^{21}$ Mx, which is most of the total unsigned flux 
($3.04\times10^{21}$ Mx) when the flux rope is initially bent.

At the photosphere, at $t = 5:11:00$, the unsigned magnetic flux reaches 
its maximum of $1.37\times10^{21}$ Mx, 45\% of the total initial flux. 
The unsigned flux at $z = 3$ Mm in the lower corona maximizes at $t = 05:29:00$ 
with a value of $8.00\times10^{20}$ Mx.
The ratio of the emerged unsigned flux 
with respect to the total initial flux decreases with increasing altitudes, 
from 86\% at $z = -3$ Mm to 26\% at $z = 3$ Mm.
The sharp decrease from the convection zone to the photosphere is caused 
by the down flows in the near-surface layers 
that tend to return flux to the deep convection zone. 

The question then, is how the horizontal and vertical flows (driven both by 
convective motions and the Lorentz force) affect the emergence 
of magnetic flux and the transport of the magnetic energy.  To address this
question we next calculate the magnetic energy flux (Poynting flux) passing 
through three layers of the atmosphere: $z = $ -3, 0, and 3 Mm using the 
following equations:

\begin{eqnarray}
  E_{\mathrm{shear}}& = &-\int \frac{1}{4\pi}\left(B_{x}u_{x} + B_{y}u_{y}\right)B_{z} dS,\\
  E_{\mathrm{emerge}}& = &\int \frac{1}{4\pi}\left(B_{x}^{2} + B_{y}^{2}\right)u_{z} dS.
\end{eqnarray}
The lower panel of Figure \ref{flux} illustrates the temporal evolution of the 
Poynting energy flux associated with the horizontal and vertical motions. 
We find that the energy flux associated with the vertical flows at $z =$ -3 and 0 Mm, 
remains negative during the rising and decaying phase of the flux emergence. 
There are several transient positive pulses of energy fluxes 
by the vertical motions at $z = -3$ Mm, at times $t = 2:09:00$ and $t = 2:57:00$. 
These transient positive pulses represent the energy transport 
when the magnetic flux first emerges to the surface with upflows, 
as shown by the lower right panel of Figure \ref{dipole}. 
Each of these pulses is followed by a sharp increase in the energy flux associated 
with the horizontal flows and a reversal of the energy flux by vertical flows. 
This time evolution of energy shows a process of magnetic flux emergence at the near 
surface layers with convective flows: magnetic flux emerges at the surface as bipoles 
with upflowing motion, then they are quickly pulled apart by the horizontal flows, 
and concentrate in the downdrafts. This process then leads to a positive energy flux
by the upflows, followed by a increasing energy flux by the horizontal motions, and 
a negative energy flux associated with the downdrafts.

\begin{figure*}[ht!]
  \begin{minipage}[t] {1.0\linewidth}
    \begin{center}
      \includegraphics[width=145mm]{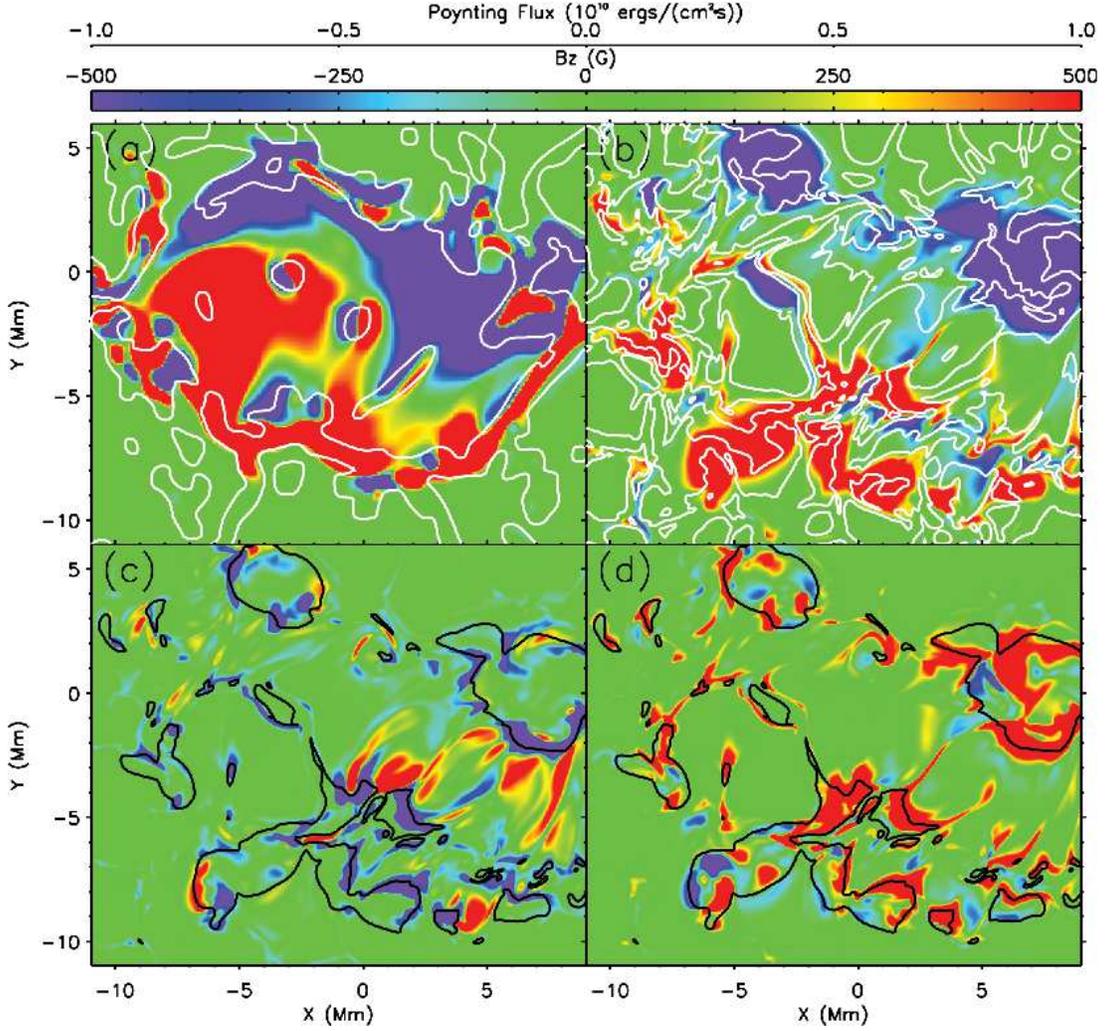}
    \end{center}
  \end{minipage}\hfill
  \caption{The structure of $B_{z}$ field at $z = -3$ Mm at $t =$ 3:40:00 (a),
    $B_{z}$ field at $z = 0$ Mm at $t =$ 4:03:00 (b) and Poynting fluxes associated 
    with vertical (c) and horizontal flows (d) at $z = 0$ Mm at $t =$ 4:03:00.
    The white lines in Panel (a) and (b) indicate downflowing regions with 
    $u_{z} =$ -0.1 km~s$^{-1}$, 
    and the black lines in Panel (c) and (d) show regions with $|B_{z}| =$ 500 G.}
  \label{bzuz}
\end{figure*}

Panel (a) and (b) in Figure \ref{bzuz} show the $B_{z}$ field at $z =$ -3 
and 0 Mm, overlaid by white lines showing downdrafts.
The concentration of the magnetic flux in the downdrafts 
explains the negative energy flux associated with the vertical flows in the lower 
panel from 3.0 hrs to 5.0 hrs while the total magnetic flux is still increasing
in the upper panel. During this time period, more than 60\% of the total unsigned 
flux is in the downflowing region for both $z =$ - 3 and 0 Mm layers.
At time $t = 5:11:00$, when the photospheric unsigned flux maximizes, 70\% of 
emerged flux is concentrated in the downdrafts. 
Panel (c) and (d) show the Poynting fluxes at the photosphere 
associated with the vertical and horizontal flows, defined as:
\begin{eqnarray}
  F_{\mathrm{shear}}& = &- \frac{1}{4\pi}\left(B_{x}u_{x} + B_{y}u_{y}\right)B_{z} ,\\
  F_{\mathrm{emerge}}& = & \frac{1}{4\pi}\left(B_{x}^{2} + B_{y}^{2}\right)u_{z} .
\end{eqnarray}
$F_{emerge}$ is negative in the magnetic polarities while postive in the areas between 
the polarities, where the magnetic flux is emerging. In the pores, the energy flow 
is instead dominated by the $F_{shear}$, shown by Panel (d). 
At $z = 3$ Mm, in the corona, the energy flux by the vertical flows shows 
short periods of postive values. The increased energy flux is driven upward by 
convectively-driven magnetoacoustic shocks, and could be interpreted in terms 
of the dynamics of Type I spicules (see e.g., \cite{hansteen2006,martinez2009}).

The Poynting flux associated with the horizontal flows dominates the energy transport 
during the flux emergence. At the photosphere, the total energy transport by this flux 
is $1.35\times10^{32}$ ergs, while vertical flows transport $5.77\times10^{31}$ ergs, 
42\% of the energy back to the convection zone by the end of the simulation. 
The horizontal flows here include the rotation of the magnetic pores, 
the separating motion of the small bipoles and the shearing motion along PILs. 
The rotation of the pores is discussed in Section \ref{rotation}, which transports 
both the magnetic energy and helicity into the corona region. The extension of 
the rotation motion into the deep convection zone shown by Figure \ref{roty=0} 
greatly impacts the spatial distribution of the Poynting flux by the rotation. 
Figure \ref{bzz=0} shows the separating process of the small bipoles, which tend 
to move apart from each other after their emergence at the photosphere and 
merge into large pores with the same polarity via this self-sorting process. 
The horizontal separating flow on the small bipoles builds up energy in the 
near surface region.  In Panel (c) and (d) of Figures \ref{bzz=0} and \ref{bzz=-3}, 
the black rectangles outline the area with a large-scale magnetic flux 
cancellation. The converging flows across the PIL build up the magnetic gradient 
in this area, and the shearing flow along the PIL, shown in Panel (c) and (d) of 
Figure \ref{bzz=0}, drives the magnetic field lines along the PIL and builds 
up a highly sheared magnetic field configuration, which plays an important role 
in eruptive events such as flares, filament eruptions and CMEs
\citep[]{mikic1994,wu1997,antiochos1999}.
A detailed analysis of this flux cancellation event will be the topic of a future paper.   

\section{Discussion and Conclusions} \label{conclusion}
In Section \ref{dipoles} and \ref{rotation}, we examined the structures of the 
vertical and horizontal velocity fields separately and their influence 
on the emerging magnetic flux rope.  The persistent large-scale downflows 
help form and maintain the bipolar pores of uniform polarity in the deep 
convection zone, while the small-scale downflows in the near-surface layers 
intensify the strength of the emerged flux tube by convective collapse, 
as suggested by \cite{parker1978} and found in previous simulations
by \cite{stein2006}.  At the photosphere, the horizontal flows act on the small 
newly-emerged bipoles, to not only separate opposite polarities, but also 
sort them into large pores of uniform polarity. This coalescence of small-scale 
fluxes enables the formation of the magnetic pores with a unsigned flux of up to 
$1.37\times10^{21}$ Mx, which is limited by the scale of our simulation. 
The horizontal flows of the magnetic pores exhibit a coherent rotation 
extending down to the deep convection zone, which is clearly driven by the 
Lorentz force as was found in earlier simulations \citep[]{longcope2000, fan2009}.  
Our simulation illustrates that this rotation mechanism operates in a 
realistic convection zone but requires the coalescence of a well formed pore 
and that complex interactions that break the symmetry of the pore, 
(such as flux cancellation) may disrupt the rotation.  

The energy transport due to the horizontal and vertical flows is calculated in 
the domain from the convection zone into the corona.
\cite{abbett2011} finds the Poynting energy flux flows into the interior below the 
visible surface and flows into the corona above it and suggests surface
convection as the energy source of the separatrix in the energy flux. 
Our study shows a negative total energy flux at $z = -3$ Mm, and positive at $z = 0$ 
and 3 Mm, which agrees with this result.  The energy flux is initially dominated 
by the emerging flow on the magnetic flux, quickly after which opposite polarities 
separate, giving rise to an increase in the energy flux associated with the horizontal 
flows. This general trend has been found in early work e.g. 
\cite{magara2003, manchester2004, fang2010}, but these works did not illustrate the
extent to which downflows control energy transport in the convection zone.  
In the convection zone and at the photosphere, the flux concentrates in the downflow 
drafts, yielding a negative energy flux associated with the vertical motion. 
The horizontal flows are thus the dominating carriers of the energy transport 
into the corona. 
We identify three types of horizontal flows in Section \ref{rotation}: the rotation, 
the separating motion of the bipoles, and the shearing motion along PILs. The rotation 
and shearing both transfer the magnetic energy and helicity into the corona. 
And it is the separating and shearing motions that build up the magnetic energy 
in the near-surface region, which is crucial for the eruptive events. 

\begin{figure*}[ht!]
  \begin{minipage}[t] {1.0\linewidth}
    \begin{center}
      \includegraphics[width=140mm]{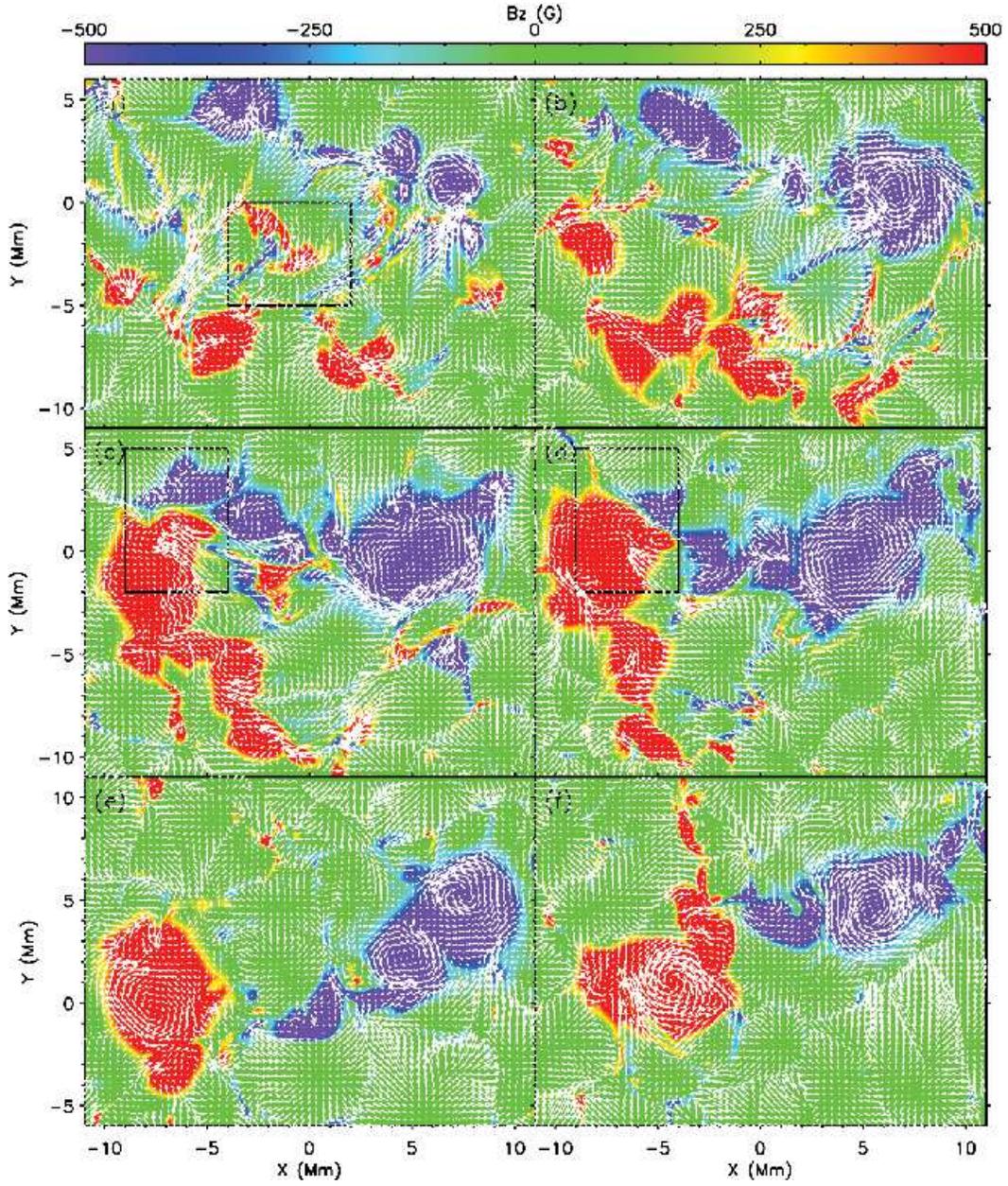}
    \end{center}
  \end{minipage}\hfill
  \caption{The structure of $B_{z}$ field at $z = 0$ Mm at t = 3:45:00 
    (a), 4:15:00 (b), 5:10:00 (c), 5:35:00 (d), 6:23:00 (e) and 7:41:00 (f). 
    The arrows show the horizontal velocity field.}
  \label{bzz=0}
\end{figure*}

\begin{figure*}[ht!]
  \begin{minipage}[t] {1.0\linewidth}
    \begin{center}
      \includegraphics[width=105mm]{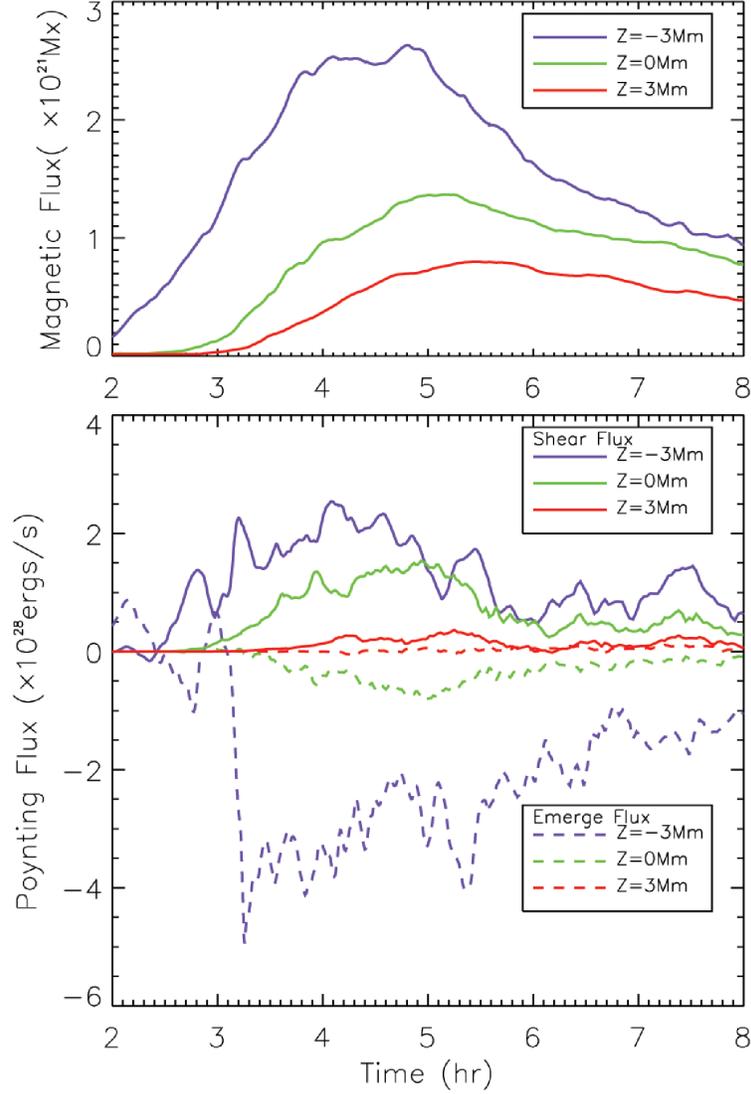}
    \end{center}
  \end{minipage}\hfill
  \caption{The temporal evolution of magnetic flux (upper) and 
    energy flux (lower) associated with the horizontal flows (solid) and
    vertical flow (dashed) at $z =  -3$ (purple), 0 (green) and 3 (red) Mm from time 
    $t = 2:00:00$ to $t = 8:00:00$. }
  \label{flux}
\end{figure*}

In our simulation, we also find an area with cancellation of magnetic flux up to 
$10^{20}$ Mx within 0.5 hour, shown by Panels (c) and (d) of Figure \ref{bzz=0}. 
The coronal response to this large-scale flux cancellation is of great interest. 
\cite{green2011} finds the evolution of coronal fields into a highly sheared arcade
and then a sigmoid in an active region where $1/3$ of the flux is cancelled. 
Whether the cancelled flux in our simulation produces a flux rope in the coronal region 
will be the topic of a future paper. Another interesting feature is the decay of the 
magnetic pores from time $t = 6:00:00$ to $t = 8:00:00$, shown by 
Figure \ref{flux}. \cite{ball2007} suggests that the submerged magnetic field repairs 
the toroidal flux ropes from which the initial flux emerged. The decayed magnetic 
flux thus may play a fundamental role in replacing the magnetic field in the solar 
interior.

\acknowledgments
This work was supported by NASA grant NNG06GD62G, NNX07AC16G and NSF grant ATM 
0642309 and AGS 1023735. W. M. IV was also funded by NASA grant LWS NNX09AJ78G.
W. P. A. was funded in part by NSF grants ATM-0641303 and AGS-1048318, and 
through the NASA Living with a Star TR\&T program (NNX08AQ30G) and the NASA 
Heliophysics Theory Program (NNX08AI56G and NNX11AJ65G).
The simulations described here were carried out on the Pleiades system at the 
NASA Advanced Supercomputing (NAS) Facility and Bluefire cluster at NCAR. 

\bibliographystyle{apj}
\bibliography{ref}

\clearpage

\end{document}